\documentclass[%
 reprint,
 amsmath,amssymb,
 aps,
]{revtex4-2}

\usepackage{graphicx}
\usepackage{dcolumn}
\usepackage{bm}
\usepackage[normalem]{ulem}

\begin{document}

\preprint{APS/123-QED}

\title{Positron Driven High-Field Terahertz Waves in Dielectric Material}

\author{N. Majernik$^1$,  G. Andonian$^1$, O. B. Williams$^1$, B. D. O'Shea$^2$, P. D. Hoang$^1$,  C. Clarke$^2$, M. J. Hogan$^2$, V. Yakimenko$^2$, J. B. Rosenzweig$^1$}

\affiliation{
$^1$UCLA Department of Physics and Astronomy, Los Angeles, California 90095, USA\\
$^2$SLAC National Accelerator Laboratory, Menlo Park, California 94025, USA
}

\date{\today}

\begin{abstract}
Advanced acceleration methods based on wakefields generated by high energy electron bunches passing through dielectric-based structures have demonstrated $>$GV/m fields, paving the first steps on a path to applications such as future compact linear colliders. 
For a collider scenario, it is desirable that, in contrast to plasmas, wakefields in dielectrics do not behave differently for positron and electron bunches.
In this Letter, we present measurements of large amplitude fields excited by positron bunches with collider-relevant parameters (energy 20 GeV, and $0.7 \times 10^{10}$ particles per bunch) in a 0.4 THz, cylindrically symmetric dielectric structure. 
Interferometric measurements of emitted coherent Cerenkov radiation permit spectral characterization of the positron-generated wakefields, which are compared to those excited by electron bunches. 
Statistical equivalence tests are incorporated to show the charge-sign invariance of the induced wakefield spectra.
Transverse effects on positron beams resulting from off-axis excitation are examined and found to be consistent with the known linear response of the DWA system. The results are supported by numerical simulations and demonstrate high-gradient wakefield excitation in dielectrics for positron beams. 
\end{abstract}

\maketitle

Relativistic positrons and electrons in high-luminosity linear colliders are essential tools for fundamental studies in particle and nuclear physics. 
The large electron positron (LEP) collider at CERN operated at up to 209~GeV center-of mass-energy \cite{Barate}, however yet higher energy enables important studies into high-precision fundamental particle mass measurements, exotic Higgs particle couplings \cite{Barklow-higgscoupling}, the search for dark matter \cite{Chae_2013}, or physics beyond the standard model \cite{Liu_2017}. 
Achieving TeV-scale center-of-mass energy for positron-electron collisions in a practical footprint \cite{ilc, clic} is a major challenge given existing radio-frequency (RF) acceleration technologies, which are limited by material breakdown to accelerating gradients of $\sim$80~MeV/m.
While cryogenic RF cavities have shown improvements to the breakdown limit \cite{Cahill}, even larger gains in accelerating electric fields are necessary to enable the next generation of TeV-class machines with reasonable cost and physical footprint.  
Advanced acceleration methods based on wakefields in a variety of media are capable of generating accelerating gradients up to three orders of magnitude greater than present RF-based accelerators, offering a path to more compact and affordable high energy physics instruments.
For example, the concept of a high gradient, wakefield-based ``afterburner'' \cite{lee2002energy,thompson2006ultra}, where the terminal energy of a relativistic beam from an existing accelerator facility is multiplied several-fold beyond the design energy of the machine,  is an attractive approach for a future $e^+$$e^-$ linear collider.

Wakefield acceleration techniques use one or more low-energy \emph{driver} bunches to excite electromagnetic fields, which can accelerate a trailing \emph{witness} bunch by extracting energy from the driver, essentially transforming a high-current, low voltage system into a relatively lower current, high voltage system. 
Plasma wakefield acceleration (PWFA) has demonstrated $>$GV/m accelerating gradients \cite{litos2014high, blumenfeld2007energy, deng2019generation} by exciting a nonlinear plasma oscillation, where the plasma electrons are evacuated from the beam channel via the repulsing space-charge of the  electron bunch driver. This leaves a positively charged ion column, resulting in a linear restoring force on the beam electrons.  Such a configuration is able to provide both strong accelerating gradients and linear focusing fields for the body of the driver, as well as for electron witness bunches. However, the nonlinear beam-plasma interaction is not symmetric for positively charged witness beams, \emph{e.g.} beams of positrons. 
Reconfiguration of the plasma profile geometry to that of a hollow-core column alleviates some issues related to positively-charged beams, and has shown experimental progress, \cite{gessner2016demonstration,doche2017acceleration}, yet stable positron acceleration in a PWFA remains an urgent challenge under study \cite{PhysRevLett.127.104801, lindstrom2018measurement}. 
Dielectric wakefield acceleration (DWA), in contrast, relies on wakefields excited by a driver bunch in a solid-state dielectric structure rather than a plasma. High gradient (GeV/m-class) acceleration of electrons has been demonstrated in dielectric wakefield accelerators \cite{oshea2016observation} but, prior to this work, only electron bunches have been used in DWA. 
In light of the plasma wakefield accelerator challenges described above, DWA has advantage for use as a positron accelerator because, in the linear limit, DWA is independent of charge-sign and can accelerate positrons with no fundamental change in the layout of the interaction. 
Experimentally observed high field damping effects in electron-driven DWA \cite{oshea2019conductivity} provide a presumed practical limit to the operational gradient; in simple cylindrical SiO$_2$ DWA geometries this is near 850 MeV/m.
Further, charge-sign specific effects, such as field emission driven by the intense electric field from a positively charged bunch, must be considered: electrons pulled from the dielectric may form an electron cloud within the vacuum volume, interfering with the beam quality, the wakefield generation and propagation, and possible beam-breakup instability due to the interaction \cite{electroncloud}.
With these considerations in mind, in this Letter, we report the first demonstration of high gradient DWA driven by a positron beam. This is an important initial step in understanding relevant effects on relativistic positron beam propagation in a collider-relevant parameter regime, a necessary milestone on the road towards development of an $e^+$$e^-$ collider based on DWA technology.


\begin{figure}[t]
   \centering
   \includegraphics*[width=\columnwidth]{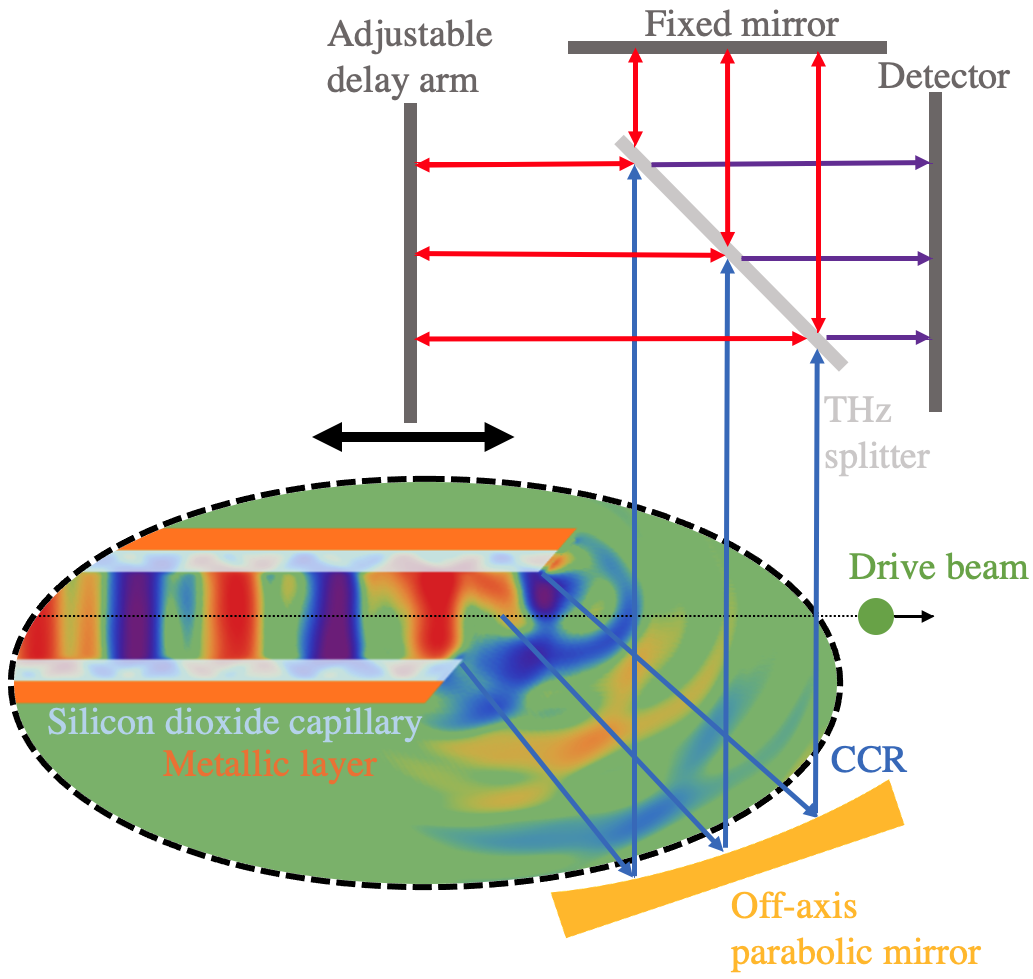}
   \caption{(Not to scale) Schematic overview of the experiment. The particle beam excites CCR in the dielectric-lined waveguide (shown with $E_z$ fields from numerical simulations) which is quasi-optically propagated to a THz-interferometer.}
   \label{fig:experimentSchematic}
\end{figure}

The issue of charge symmetry in DWA is examined in a set of experiments conducted at the Facility for Advanced Accelerator Experimental Tests (FACET) at SLAC National Accelerator Laboratory \cite{Hogan:2010}.
The facility employs high-energy electron and positron bunch modalities enabling a variety of studies. 
In the positron bunch DWA experiment, accessible properties of the excited wakefields are characterized, and compared to those excited by electron bunches. 
When establishing the positron bunch parameters, it is important as noted above to operate at accelerating gradients below the threshold for onset of high-field damping effects in dielectrics \cite{oshea2019conductivity}. Also, one should employ a suitable dielectric media, to avoid potential field emission from the bulk material due to the high fields generated by the positron beam.
Finally, demonstrating equal charge-sign response requires employing statistical equivalence tests in data analysis, because the positron and electron data sets were acquired at different times, under subtly differing systematic conditions.

At FACET, positron beams are generated by colliding a 20 GeV electron beam onto a tungsten alloy target, creating e$^+$/e$^-$ pairs. The positrons are then collected, cooled, and transported to the interaction region \cite{SLACbook}.
For the set of experiments, for both electron and positron drive bunches, the initial energy was 20 GeV, with $N_b=0.7 \times 10^{10}$ particles per bunch, and $\sigma_x = \sigma_y = \sigma_z = 40 \; \mu$m. The DWA structure consists of silicon dioxide (SiO$_2$), a nonpolar material, in a cylindrical geometry. The SiO$_2$ capillary is coated on the outer wall with a $\sim$10~$\mu$m thick copper layer. 
The inner and outer radii of the dielectric structures are approximately 200 $\mu$m and 300 $\mu$m respectively (\textit{i.e.}  having a 100 $\mu$m wall thickness), with an overall length of 3~cm.  
As the beam transits the dielectric structure, coherent Cerenkov radiation (CCR) is excited within the dielectric; for the parameters of this experiment, the CCR is in the terahertz band. 
The radiation propagates forward and is launched from the end of the tube, which is impedance matched to free space with a Vlasov-type antenna \cite{Vlasov:1974}. The CCR is propagated quasi-optically using a short-focal length off-axis parabolic mirror to the main diagnostic, a THz-sensitive interferometer (See Figure \ref{fig:experimentSchematic} for a schematic overview of the experiment). 
An autocorrelation-based interferogram is produced on a multi-shot basis by scanning the delay arm. 
Pyroelectric detectors, coated for enhanced sensitivity in the THz spectral range, are used in the interferometer. 
The interferometer signal is processed to calculate the spectral content of the radiation, which aids in revealing the presence of high-field damping \cite{oshea2016observation, oshea2019conductivity} or additional modes excited by off-axis propagation.

\begin{figure}[htb]
   \centering
   \includegraphics*[width=\columnwidth]{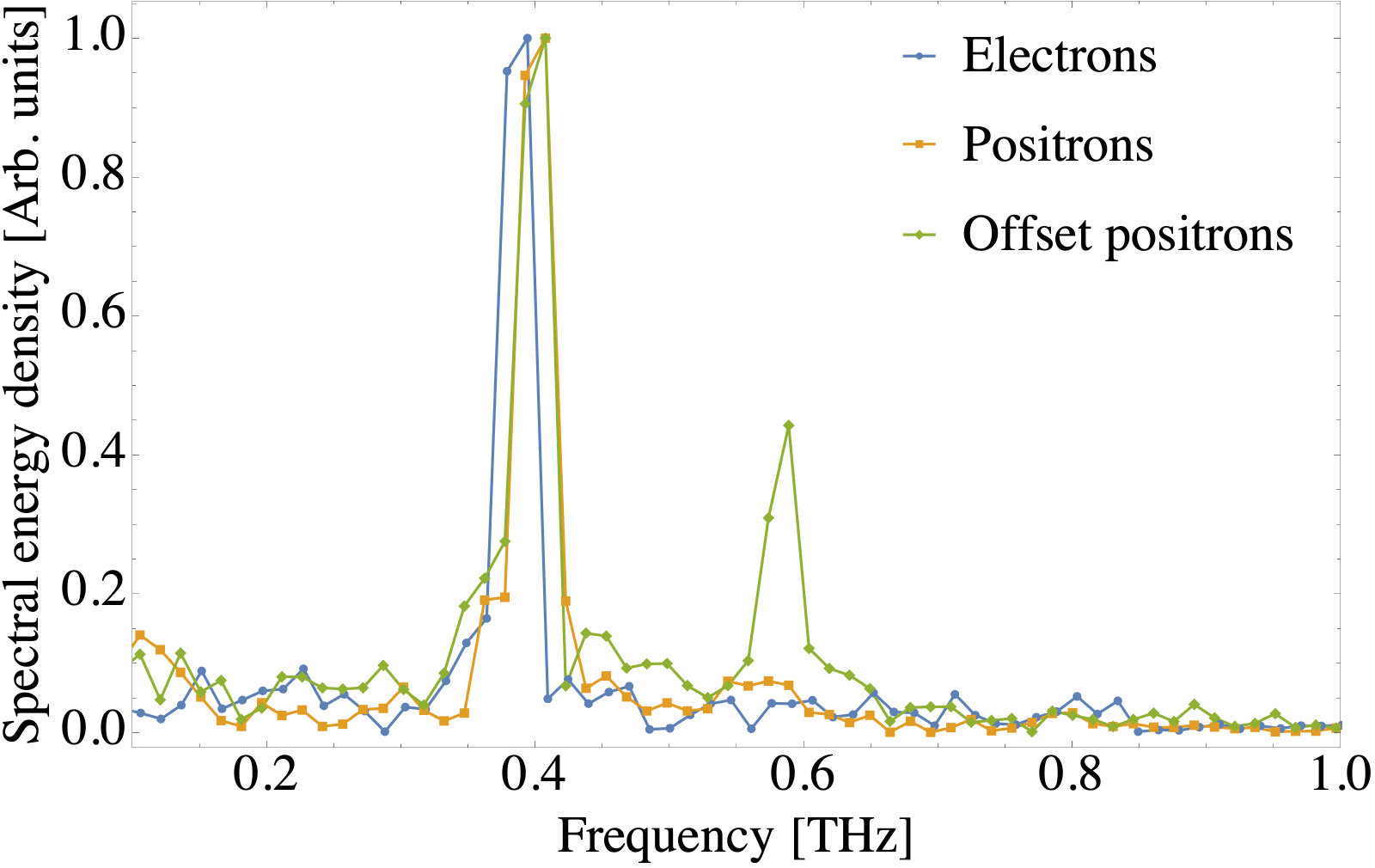}
   \caption{Average CCR spectra for electrons and positrons on-axis, as well as positrons 60 $\mu$m off-axis. The peak in all cases has been normalized to one.}
   \label{fig:FFT_measured}
\end{figure}

\begin{figure*}[t]
    \centering
    \includegraphics[width=0.8\textwidth]{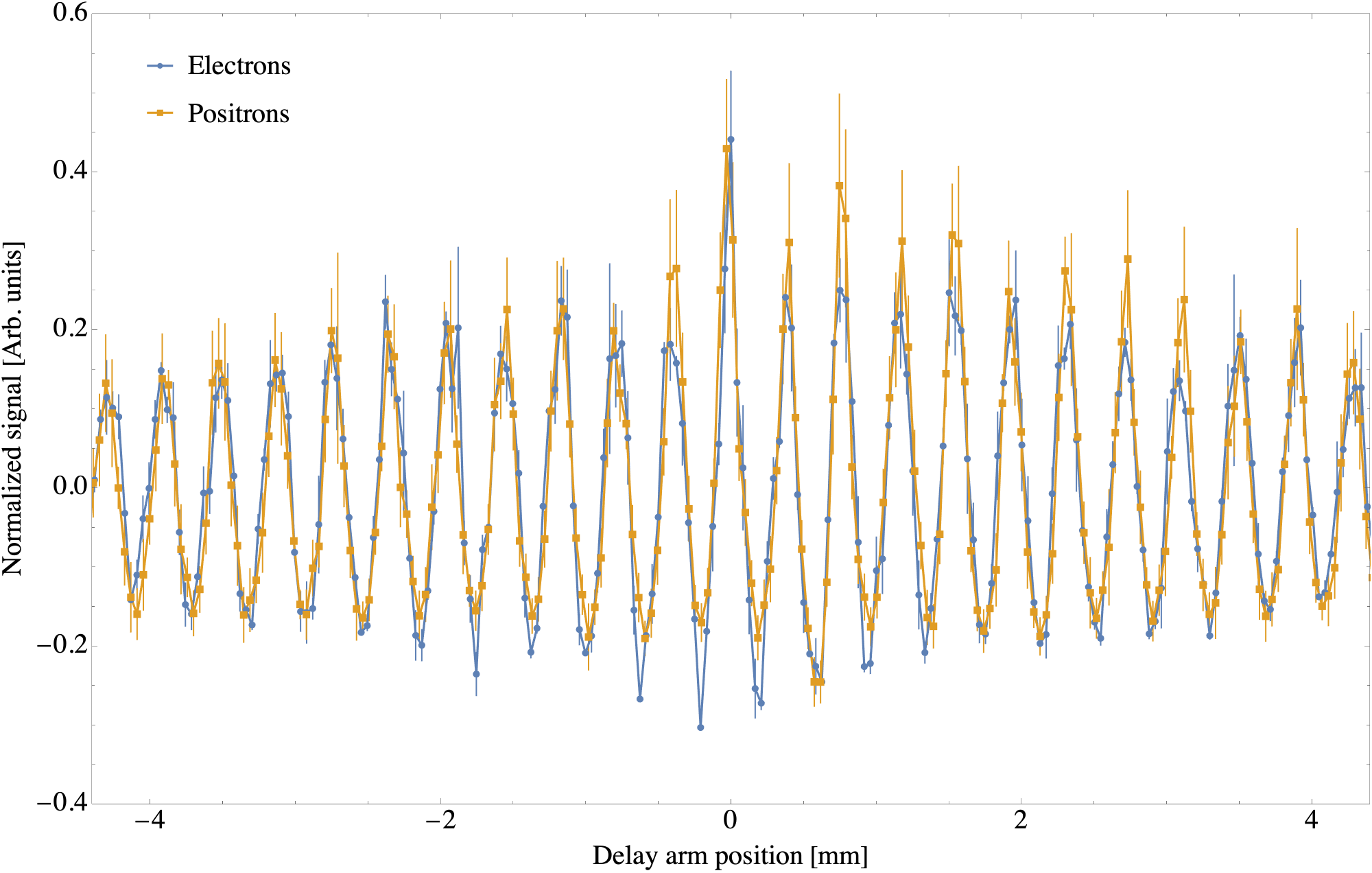}
    \caption{Frequency scaled and amplitude normalized interferograms for on-axis electrons and positrons, shown with one standard deviation error bars. The connecting lines show the linear interpolations, $m_{\{e,p\}}(z)$, employed by the statistical analysis.} 
    \label{fig:interferograms}
\end{figure*}

For a dielectric lined waveguide with cylindrical geometry, the TM-mode (accelerating) frequencies are given by the solutions that satisfy the dispersion relation \cite{zhang1998cerenkov}:
\begin{equation}
    \frac{I_1(k_1 a)}{I_0(k_1 a)} = \frac{\epsilon_r k_1}{k_2} \frac{J_0(k_2 b) Y_1(k_2 a) - J_1 (k_2 a) Y_0(k_2 b)}{J_0(k_2 b) Y_0(k_2 a) - J_0 (k_2 a) Y_0(k_2 b)}
    \label{eq:dispersionEq}
\end{equation}
where $k_1$ and $k_2$ are the radial wavenumbers in the vacuum and dielectric respectively, $\epsilon_r$ is the dielectric constant (3.8 for SiO$_2$), $a$ and $b$ are the inner and outer radii of the tube, and $I_n$, $J_n$, $Y_n$ are Bessel functions. 
The solutions to Equation~\ref{eq:dispersionEq} describe the allowed longitudinal modes in the structure of a given geometry. The convolution of these functions with the temporal profile of the driver beam yields the excited wakefields.
For the scenario described, this expression estimates that the fundamental frequency, TM$_{01}$, is 393~GHz.

The CCR autocorrelation data were collected by the interferometer, in steps of 42 $\mu$m, recording several shots per delay. The resulting data set was Fourier transformed to yield the spectra shown in Figure \ref{fig:FFT_measured}. 
Fitting the fundamental mode peaks for the electron and positron runs gives estimates for the structures' fundamental modes of 387 and 400 GHz respectively, in good agreement with the analytic value. Slightly differing cylindrical structures were used for the positron and electron data sets and the frequency discrepancy in measurements is attributable to a spread in tube dimensions consistent with manufacturing tolerances; a 4 $\mu$m difference in radius, within the manufacturing tolerance, yields 13 GHz variation in frequency.
To permit comparative analysis, the autocorrelation data for the on-axis electron and positron runs were scaled to compensate for the observed frequency difference and then normalized in amplitude and offset to minimize mean-average-error with the results shown in Figure \ref{fig:interferograms}.


\begin{figure}[b]
   \centering
   \includegraphics*[width=\columnwidth]{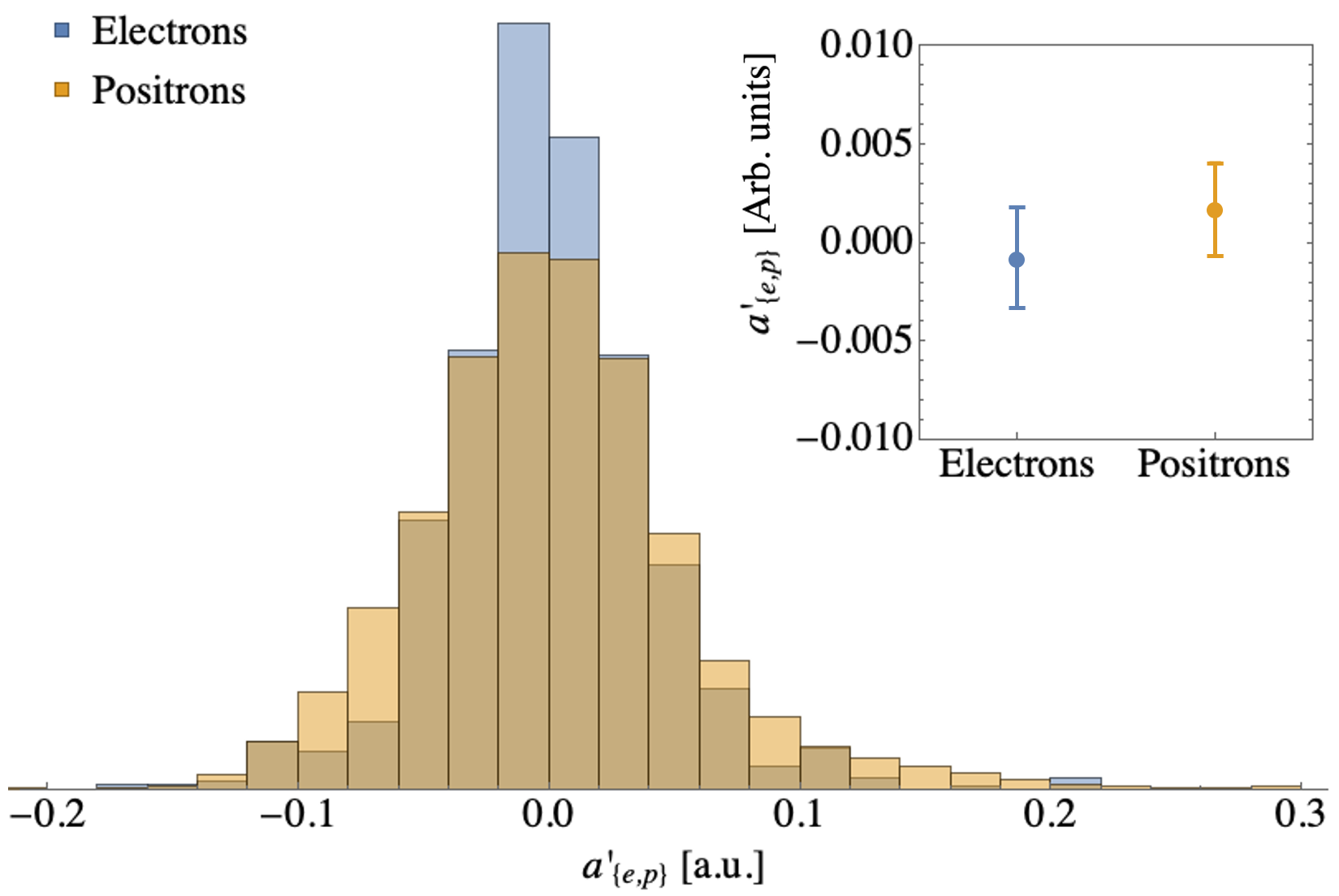}
   \caption{\textbf{Main} Histogram showing the data from Figure \ref{fig:interferograms}, transformed according to Equation \ref{eq:transform}. Vertical axis is scaled for each population according to the total number of samples. \textbf{Inset} Mean values of these transformed points with 95\% confidence interval bars.}
   \label{fig:stats}
\end{figure}

\begin{figure*}[htb]
    \centering
    \includegraphics[width=\textwidth]{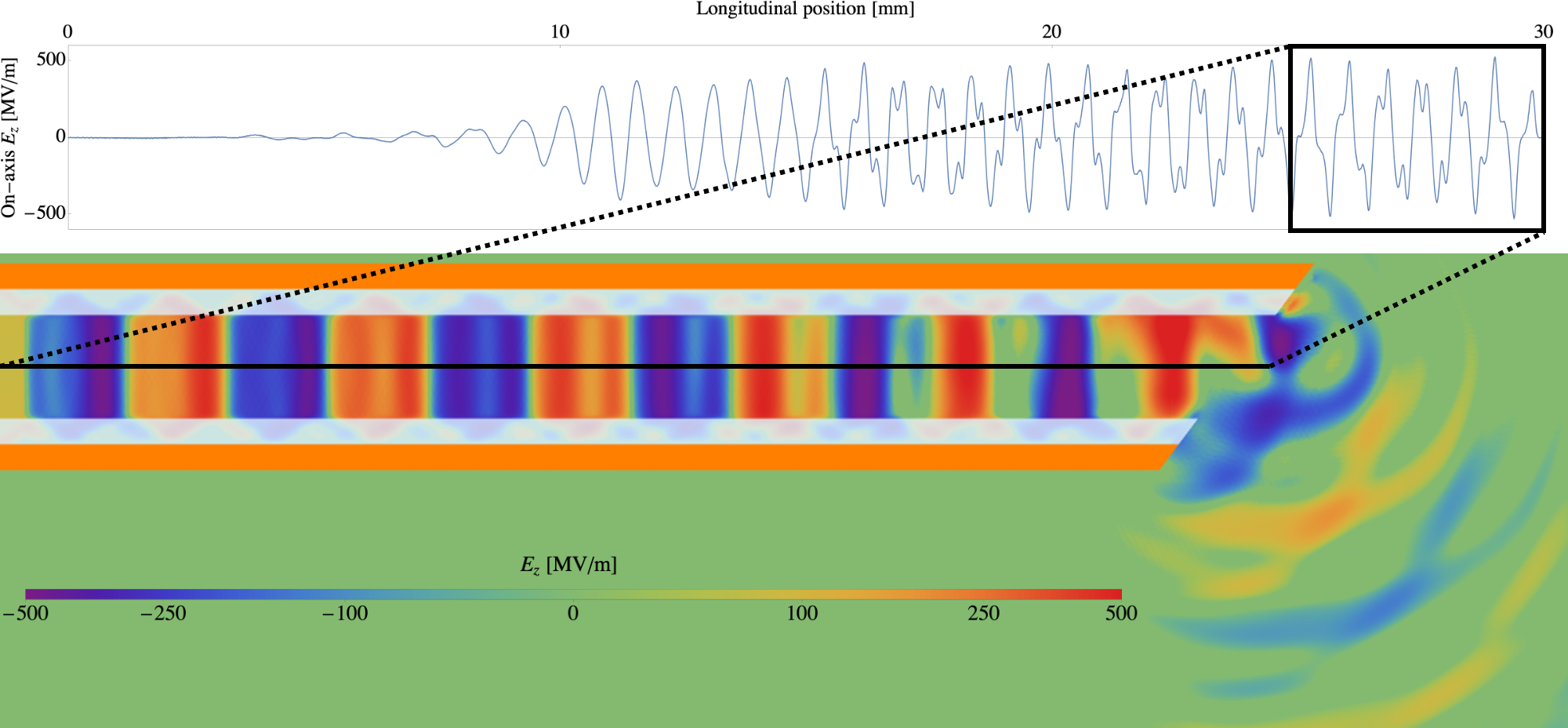}
    \caption{PIC simulation of on-axis DWA. \textbf{Top} Longitudinal electric field, $E_z$, measured on-axis along the length of the dielectric structure. The excitation of higher modes leads to a higher peak value than predicted by Equation \ref{eq:peakEz}. \textbf{Bottom} Cross section of the last 5 millimeters of the dielectric structure showing the $E_z$ fields, illustrating how the CCR excited inside the structure is coupled out into free space for measurement.} 
    \label{fig:CSTresults}
\end{figure*}

One effect that should be considered is the onset of high-field damping in SiO$_2$ DWA structures. In previous studies, it was discovered that the damping has a threshold value ($|E_z| \approx 850$ MV/m \cite{oshea2019conductivity}) for electron drivers. The peak longitudinal field in a DWA driven by a Gaussian beam can be estimated by \cite{thompson2008breakdown}:
\begin{equation}
    E_z \approx \left | \frac{4 N_b r_e m_e c^2}{a q_e \left ( \sqrt{\frac{8\pi}{\epsilon_r-1}} \epsilon_r \sigma_z + a \right )} \right |
\label{eq:peakEz}
\end{equation}
where $N_b$ is the number of particles in the drive bunch, $r_e$ is the classical electron radius, $m_e$ is the electron mass, and $q_e$ is the electron charge. For these measurements, this estimate yields $300$ MV/m. However, simulations reveal that higher order modes are being excited, increasing the peak field to $500$ MV/m (See Figure~\ref{fig:CSTresults}). 
This value is chosen because it is comfortably below the threshold for the onset of high field damping, yet still quite significant for high-gradient acceleration; this gradient is indeed comparable to recent experiments on positron plasma wakefield acceleration \cite{gessner2016demonstration}.
The interferograms in Figure~\ref{fig:interferograms} agree with expected results and 
do not display the characteristic, strong decay signature of induced-conductivity enabled damping. 
Furthermore, as evident in Figure~\ref{fig:interferograms}, the interferograms generated by positrons compared to those generated by electrons, do not reveal evidence of charge-dependent effects. 
The statistical equivalence of the interferogram data is rigorously demonstrated below. 

The electron and positron data can be demonstrated to be statistically equivalent but, due to the nature of the data and the fact that different structures were used at different times, direct application of typical statistical analytic approaches, such as paired tests, is not appropriate. 
Instead, a two one-sided equivalence test is performed on transformed populations.
For each delay arm position, multiple shots are recorded (five for the electron data set, ten for the positron data set) but these delay arm positions are not the same between data sets. 
The interferometer reading from each of these shots is denoted as $a_{\{e,p\},k}(z_i)$ where $e$ or $p$ indicates the species as either electrons or positrons respectively, $k$ is the shot number at that position, and $z_i$ is the physical location of the delay arm for the $i^\mathrm{th}$ delay arm position. For each species, at each position, we calculate the median of these values; a linear interpolation of these medians in $z$ yields two continuous functions with respect to delay arm position: $m_{\{e,p\}}(z)$. The mean trend line is defined $m_t(z) \equiv (1/2)(m_e(z)+m_p(z))$. The trend line is subtracted from each measurement creating two new populations:
\begin{equation}
a'_{\{e,p\},k}(z_i) = a_{\{e,p\},k}(z_i) - m_t(z_i),
\label{eq:transform}
\end{equation}
shown in Figure \ref{fig:stats}. By applying a two one-sided test \cite{hauck1984new,lakens2018equivalence} to these transformed populations, we can demonstrate the equivalence of the electron and positron responses. To this end, we assert that the smallest effect size of interest (SESOI) is 0.05 (in the units of Figure \ref{fig:interferograms}), approximately equal to the \mbox{per-position} standard deviation for both species. The null hypothesis, $H_0$, is that of non-equivalence to a meaningful extent, i.e. $|\mu_e-\mu_p| \ge \mathrm{SESOI}$, where $\mu_e$ and $\mu_p$ are the means of the transformed populations $a'_e$ and $a'_p$ respectively while the alternative hypothesis, $H_A$, is that of effective equivalence, $|\mu_e-\mu_p| < \mathrm{SESOI}$. At the 95\% confidence level ($p$ = 0.000) we may reject the null hypothesis and conclude that the electron and positron responses are functionally equivalent. 

\begin{figure}[h]
   \centering
   \includegraphics*[width=\columnwidth]{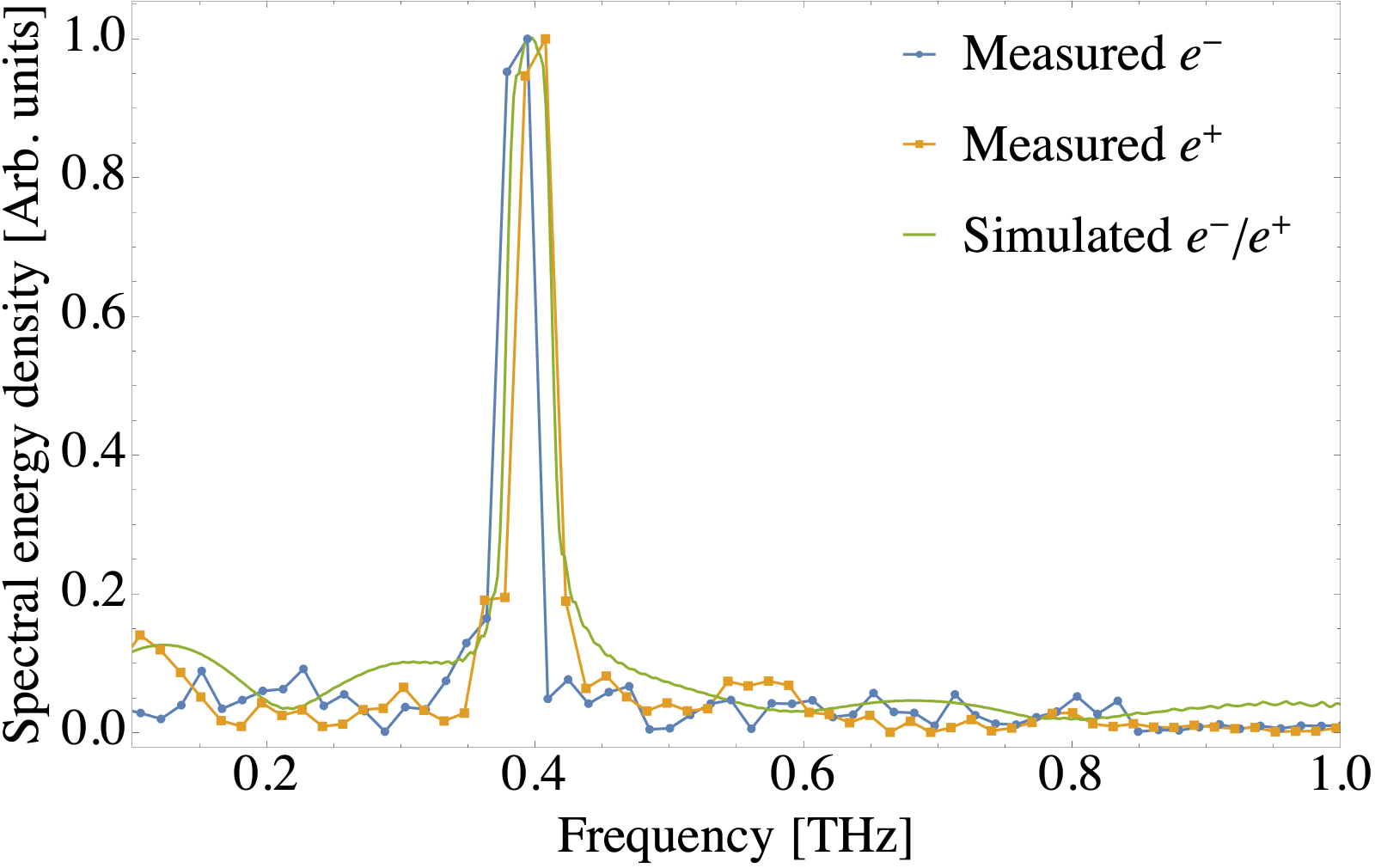}
   \caption{Average CCR spectra for electrons and positrons on-axis, shown with the spectrum of a PIC simulation of this scenario. No charge sign symmetry breaking effects were included in the simulation so the results are the same for both electrons and positrons. The peak in all cases has been normalized to one.}
   \label{fig:FFT_CST}
\end{figure}

This experimental scenario was also simulated using a particle-in-cell (PIC) code, CST \cite{CST}. Figure \ref{fig:CSTresults} shows the on-axis longitudinal field, $E_z$, and illustrates that higher order modes are excited, with peak fields reaching, as noted above, $500$ MV/m. The nature of the Vlasov antenna CCR outcoupling is also illustrated with the radiation propagating quasi-optically to the interferometer diagnostic. The spectrum of these free space fields is calculated and shown in Figure \ref{fig:FFT_CST}; the TM$_{01}$ frequency agrees with the analytic estimate of Equation \ref{eq:dispersionEq} and the measured positron and electron spectra.

In addition to acceleration from longitudinal fields, high frequency dielectric structures also generate commensurately large transverse fields \cite{bettoni2016temporal,oshea2020suppression}.
For the nominal structure dimensions, the HEM\textsubscript{12} mode is, for linear response, expected to have a frequency of 569 GHz, consistent with the measured value for a positron beam propagated 60 $\mu$m off-axis (Figure \ref{fig:FFT_measured}) of 584 GHz. As with the TM modes, this difference is attributed to the manufacturing tolerances of the capillaries. 
The transverse kick on the positron beam is also measured with downstream beam position monitors and the results are consistent with off-axis PIC simulations. 

These results provide key experimental support for the proposed application of DWA in a future e$^+$e$^-$ collider operated at high accelerating gradient, in a regime relevant to current designs \cite{Shiltsev:2021}, where head-on interactions with witness beams in the 0.8-3.2 nC range are conceptualized to achieve high luminosity.
The positron-driven wakefields described in this Letter are analogous to those induced by the positron witness beams in such a collider and our results provide evidence that long-range acceleration may be achieved without disruption by charge-sign specific, higher order effects.
We demonstrate positron-driven DWA fields behave in an equivalent fashion to electron-driven DWA fields, as confirmed by the modal analysis, for gradients up to 500 MeV/m and with collider-relevant bunch size and charge, in a relatively simple accelerating medium compared to formation of sophisticated plasma configurations for PWFA \cite{gessner2016demonstration,PhysRevLett.127.104801,doche2017acceleration}. 

With these results in hand, a scenario where both electrons and positrons can attain high final energy before interaction, with similar dielectric-lined structures, may be enabled. 
This approach can reduce the final footprint of the collider accelerators by one order of magnitude beyond the current state-of-the-art. 
Further, hybrid schemes including combining PWFA and DWA to accelerate electrons and positrons respectively, can also be envisioned, where a reduction in total acceleration on the positron side can be compensated by ultra-high field electron acceleration in plasma, to achieve the same  energy beams at the interaction. 

Continued experimental investigations on positron-driven DWA are planned to explore the possible onset of higher order, charge-sign dependent effects. 
Upcoming experiments at FACET-II \cite{Yakimenko:2019}  aim to excite and characterize  such effects by driving higher strength wakefields in smaller structures, and exploring different media responses. These effects may include anomalous damping \cite{oshea2019conductivity} or the emission of electrons from the dielectric surface into the DWA vacuum channel. These dark-current electrons are pulled from the dielectric by strong, radial fields (instead of by ionization of residual gas), and may subsequently induce an electron-cloud-like interaction \cite{electroncloud}. 
We note that high field damping and electron-cloud formation may be mitigated by the introduction of advanced dielectric structures \cite{Andonian:2014,Hoang:2018}, high bandgap materials \cite{naranjo2012stable}, or cryogenic cooling. 

We thank Dr. Fred J. Hickernell of the Illinois Institute of Technology for valuable discussions on statistical analysis. This work was supported by the Department of Energy High Energy Physics Grant DE-SC0009914.

\bibliography{References.bib}

\end{document}